\documentclass[10pt,conference]{IEEEtran}
\IEEEoverridecommandlockouts
\usepackage{graphicx} 
\usepackage{float} 
\usepackage{algorithm}
\usepackage{algorithmic}
\usepackage{color}

\usepackage{hyperref}

\usepackage[ruled,vlined,algo2e]{algorithm2e}
\usepackage{amsmath}

\title{CodeWatcher: IDE Telemetry Data Extraction Tool for Understanding Coding Interactions with LLMs}

\author{
Manaal Basha\textsuperscript{1}, 
Aimeê M. Ribeiro\textsuperscript{2}, 
Jeena Javahar\textsuperscript{1},  
Cleidson R. B. de Souza\textsuperscript{2} \\
Gema Rodríguez-Pérez\textsuperscript{1},
\textsuperscript{1}University of British Columbia, Kelowna \quad
\textsuperscript{2}Federal University of Pará \\
\texttt{manaals@student.ubc.ca}, \texttt{aimee@ufpa.br}, \texttt{jeenajavahar@gmail.com}, \\
\texttt{cleidson.desouza@acm.org}, \texttt{gerope@mail.ubc.ca}
}
\date{May 2025}

\begin{document}

\maketitle

\begin{abstract}
Understanding how developers interact with code generation tools (CGTs) requires detailed, real-time data on programming behavior which is often difficult to collect without disrupting workflow. We present \textit{CodeWatcher}, a lightweight, unobtrusive client-server system designed to capture fine-grained interaction events from within the Visual Studio Code (VS Code) editor. \textit{CodeWatcher} logs semantically meaningful events such as insertions made by CGTs, deletions, copy-paste actions, and focus shifts, enabling continuous monitoring of developer activity without modifying user workflows. The system comprises a VS Code plugin, a Python-based RESTful API, and a MongoDB backend, all containerized for scalability and ease of deployment. By structuring and timestamping each event, \textit{CodeWatcher} enables post-hoc reconstruction of coding sessions and facilitates rich behavioral analyses, including how and when CGTs are used during development. This infrastructure is crucial for supporting research on responsible AI, developer productivity, and the human-centered evaluation of CGTs. Please find the demo, diagrams, and tool \href{https://osf.io/j2kru/files/osfstorage}{here}.
\end{abstract}

\begin{IEEEkeywords}
Code Generation, Telemetry, VS Code, Plugin 
\end{IEEEkeywords}

\section{Introduction}

Understanding how programmers write code, beyond just a correct final product has been a challenge in both computer science education and in industry~\cite{barke2023grounded}. This understanding is essential for improving tool support, enhancing productivity, and identifying usability challenges within IDEs. 
Traditional methods of evaluating code such as test cases or static analysis typically checks correctness or style after the code is written. These approaches although helpful, offer limited insight into the process of writing code and can struggle to support developers during moments of confusion, experimentation, or problem solving~\cite{edwards2004testdriven}.

In educational contexts, novice programmers frequently struggle to identify and correct mistakes as they write code, particularly when using code generation tools (CGTs)~\cite{barke2023grounded} such as GitHub Copilot\footnote{https://github.com/features/copilot} or Amazon Q\footnote{https://aws.amazon.com/q/} which generate code within the IDE. Providing timely, formative feedback during the coding process can improve learning outcomes, retention, and engagement in programming~\cite{shen2019, veerasamy2022formative}. However, large class sizes or online settings often make it infeasible for instructors to deliver personalized feedback at scale. Existing tools such as AutoGrader\footnote{https://autograder.io/}, Coding Rooms\footnote{https://www.codingrooms.com/}, and CodeWorkout\footnote{https://codeworkout.cs.vt.edu/} automate assessments, but typically focus on correctness or syntax, missing deeper behavioural indicators like confusion, excessive rewriting, or blind copy-pasting. While process-aware tools such as OverCode~\cite{overcode2015} attempt to visualize student code patterns, they often lack real-time monitoring and rely on batch analysis.

In industry, software engineering teams also benefit from understanding how developers code in relation to tasks such as during onboarding~\cite{ju2021case}, debugging~\cite{hirsch2021we}, and code reviews~\cite{turzo2024makes}. Prior research has shown that developer telemetry, which is data about interactions with editors, compilers, and version control systems, can identify productivity bottlenecks and support developer success~\cite{beller2025s, cui2024productivity, zeigler2024p, meyer2014}. However, most industrial tooling is proprietary, emphasizes aggregate metrics, and lacks detailed or customizable feedback suitable for educational use or behavioral analysis. Furthermore, current CGTs are based on LLMs that offer single and multiline suggestions and have been affecting developers' workflow. More studies about these tools rely on surveys, video recordings, and interviews~\cite{groundedCopilot, vaithilingam2022expectation, xu2022ide, Mendes2024, trustInAIRESEARCHPAPER} (with two exceptions~\cite{zeigler2024p, tang2024codegrits}). These studies primarily focus on users' reactions rather than capturing real-time CGT interaction data within the IDE, an area where \textit{CodeWatcher} offers a unique contribution. 


\textit{CodeWatcher} addresses the lack of detailed and scalable interaction telemetry data for CGTs used in the IDE. \textit{CodeWatcher} is a client-server system that captures detailed programming events (e.g., insertions, deletions, copy/paste, focus shifts), is compatible across multiple CGTs and programming languages, and can support real-time feedback through a rule-based engine.  Instructors or team leads can define custom rules that trigger alerts or suggestions based on coding behavior data. Unlike traditional tools that focus solely on code correctness, \textit{CodeWatcher} highlights how the code was written with CGTs, offering process-level visibility like frequent rework, use of LLMs, or idle time.


\section{Related Work}

That development of tools that capture and analyze user interactions within IDEs is crucial for understanding programmer behaviour and improving software development processes. Existing research highlights the need for tools that can track various aspects of developer activity, including code changes, IDE commands, and cognitive processes. For instance, TaskTracker, a plugin for IntelliJ-based IDEs, collects code snapshots and user actions during programming tasks \cite{lyulina2021tasktracker}. Similarly, DFLOW records interaction data, enabling analysis of time spent on different activities like editing and navigating code. These tools demonstrate the value of tracking IDE interactions for understanding developer behavior. However, they often lack the ability to simultaneously capture other forms of behavioral data, such as eye-tracking data, which can provide insights into cognitive processes~\cite{minelli2015}. 

To address this limitation, CodeGRITS was developed as a plugin for JetBrains IDEs to track both IDE interactions and eye gaze data~\cite{tang2024codegrits}. This tool allows researchers to combine these complementary approaches for a more comprehensive understanding of developer behavior. Furthermore, the emergence of LLMs in software development necessitates tools that can track the use and impact of AI-generated code. For instance, WaitGPT transforms AI-generated code into an interactive visual representation, enabling users to monitor and steer data analysis performed by LLMs~\cite{xie2024waitgpt}.

Techniques to assist with distinguishing between human-written and CGT code across multiple coding languages with high accuracy such as through AI code stylometry classifiers \cite{gurioli2024you} could help to better understand user interactions in CGTs. However, this method can be tedious as it requires having and utilizing a relevant dataset to train a transformer based encoder classifier for such a task. 

While these tools or methods offer valuable insights into developer behavior and the use of LLMs, a gap remains in tools that can specifically track data or metrics related to AI code suggestions, acceptance rates, and code reuse patterns. The \textit{CodeWatcher} tool aims to fill this gap by providing researchers with the ability to extract and analyze data related to AI-generated code within the IDE, enabling a deeper understanding of the impact of AI on software development practices.

\section{Event Capture and Representation}

A core capability of \textit{CodeWatcher} lies in its ability to unobtrusively capture detailed programmer activity by logging semantically meaningful events from the developer's IDE. It specifically logs user interactions with CGTs and not general coding behavior. The system is designed to balance fidelity with minimal intrusiveness, enabling continuous, real-time data extraction of code authoring behaviors without requiring changes to user workflows. 

The \textit{CodeWatcher} client plugin monitors eight distinct event types that reflect key aspects of interaction during a coding session: \textbf{Start}, \textbf{End}, \textbf{Insertion}, \textbf{Deletion}, \textbf{Focus}, \textbf{Unfocus}, \textbf{Copy}, and \textbf{Paste}. Each event is recorded as a structured object containing up to four fields:

\begin{itemize}
    \item \textbf{Type} – the event category (e.g., \textit{Insertion}, \textit{Focus});
    \item \textbf{Time} – a timestamp (ms precision) marking the event occurrence;
    \item \textbf{Text} – the relevant code snippet or user input (when applicable); and
    \item \textbf{Line} – the full line of code where the event occurred (for insertions and deletions).
\end{itemize}

Not all fields are applicable to every event type. Table~\ref{tab:event-types} summarizes the semantics of each event and the associated fields.

\begin{table}[h!]
\center
\footnotesize
\caption{Event Types and Captured Fields in CodeWatcher}
\label{tab:event-types}
\begin{tabular}{|p{1cm}|p{3cm}|p{3.4cm}|}
\hline
\textbf{Event Type} & \textbf{Description} & \textbf{Captured Fields} \\
\hline
\textbf{Start} & Triggered when a new file is opened and becomes active in the editor. & \textbf{Time} - records the moment the document is opened and editing begins \\ 
\hline
\textbf{End} & Triggered when the file is removed from focus. & \textbf{Time} - captures the moment the editing session ends. \\ 
\hline
\textbf{Insertion} & Logged when code is pasted from outside the IDE (including IDE chats) or code suggestion acceptances that are more than 3 characters, reducing noise from keystroke input. & \textbf{Time} - captures the time when the increase in text is detected, 

\textbf{Text} - the newly inserted text, 

\textbf{Line} - the entire line of code where the insertion occurred \\ 
\hline
\textbf{Deletion} & Logged when any content is removed, whether single-character or multi-line. & \textbf{Time} - the time when the text is deleted, 

\textbf{Text} - the text of the removed content, 

\textbf{Line} - the line where the deletion took place \\ 
\hline
\textbf{Unfocus} & Recorded when the user switches away from the IDE. & \textbf{Time} - records the time when the IDE loses focus\\ 
\hline
\textbf{Focus} & Recorded when the IDE regains focus from another application. & \textbf{Time} - records the time when the IDE regains focus\\ 
\hline
\textbf{Copy} & Detected upon executing a copy command (e.g., \texttt{CTRL+C}) within the IDE. & \textbf{Time} - the time when the copy command is executed, 

\textbf{Text} - the copied content \\ 
\hline
\textbf{Paste} & Triggered on paste command (e.g., \texttt{CTRL+V}); the subsequent \textit{Insertion} captures the actual content. & \textbf{Time} - the time when the paste command is executed 

\textbf{Text} - the newly pasted text \\
\hline
\end{tabular}
\end{table}

The plugin is optimized for minimal overhead, relying on native editor hooks to track text manipulation and focus changes. To reduce noise, it filters out character-level inputs and aggregates small edits into higher-level \textit{Insertion} and \textit{Deletion} events. Importantly, the system disambiguates between typed input, specifically collecting code generation acceptance data and pasted content.
By capturing this rich sequence of temporally ordered, context-aware events, \textit{CodeWatcher} enables both real-time intervention and post-hoc reconstruction of development sessions involving CGTs. This data forms the basis for CGT usage analysis, behavioral modeling, and personalized feedback mechanisms.

\section{Tool Architecture}

\textit{CodeWatcher} is implemented as a modular, client-server system composed of three primary components: a VS Code plugin (client), a RESTful API (server), and a backend database. This architecture supports ease of maintenance, deployment scalability, and flexible integration with workflows. 

\subsection{Client-Server Communication}

The front-end component of \textit{CodeWatcher} is a VS Code plugin written in JavaScript. Operating locally within the developer's environment, the plugin is responsible for unobtrusively capturing interaction events within the editor such as code insertions, deletions, and focus changes. This is then transmitted to the backend API over HTTP when there is an \textit{end} event type. 

Communication between the plugin and the server adheres to RESTful principles. Event data is encoded in JSON format and sent via HTTP requests to predefined API endpoints. The API, implemented in Python using the FastAPI framework, is designed around the service-repository architectural patterns to promote separation of concerns. 

The API exposes a variety of endpoints that support key operations including user account creation, login, interaction log registration, user deletion, and permission management. Input validation is handled via Pydantic models, ensuring data consistency and integrity prior to storage. All communication is routed through a remote server running Ubuntu, where the API and the database are deployed using Docker containers.

\subsection{Database Schema}

The backend database is implemented using MongoDB, selected for its schema-less structure and flexibility in handling diverse interaction log formats. Two primary collections are part of the storage layer: \texttt{user}, and \texttt{interaction\_logs}. The \texttt{user} collection stores all user credentials, metadata, and access-level permissions. Whereas \texttt{interaction\_logs} contain structured records of all IDE events, each annotated with metadata like those seen in Table \ref{tab:event-types}.

Event documents are stored as JSON-like objects, enabling dynamic schema evolution as the set of tracked events expands. MongoDB Compass is used for monitoring and querying the database during development and analysis of phases. 

\section{Validation and Testing}

To validate \textit{CodeWatcher}, we confirmed the accuracy and completeness of the data it collects and transmits to the server. This includes ensuring the tool captures the correct event types, attributes, and timestamps during development activity.

\subsection{Validation Methodology and Results}

We validated \textit{CodeWatcher}'s ability to accurately and reliably log user interactions by designing a set of controlled test scenarios (Table~\ref{tab:validation_metrics}) aligned with the following key objectives:

\begin{itemize}
  \item \textbf{Correct Event Detection:} Each user interaction (e.g., insertions, deletions, copy-paste, file start/end) is correctly identified and logged.
  \item \textbf{Field-Level Accuracy:} Captured metadata (e.g., time, text, line content) precisely reflects the editor state.
  \item \textbf{Transmission Integrity:} Ensures events are transmitted to the server without data loss or corruption.
  \item \textbf{System Robustness:} Operates reliably under realistic and edge-case IDE usage patterns.
\end{itemize}

To verify these goals, we ran 150 trials with 2 users simulating realistic developer workflows. For each event type, we triggered interactions manually and checked server-side logs for accuracy. Our validation process used:


\begin{itemize}
  \item \textbf{Exact Match Verification:} For event types involving code changes (e.g., Insertion, Deletion, Paste), we confirmed that the logged \texttt{Text} and \texttt{Line} fields exactly matched the actual editor content at the time of the event.
  \item \textbf{Timestamp Validation:} We ensured that all logged events reflected the correct temporal order by comparing timestamps to externally recorded wall-clock timings (e.g., from screen recordings).
\end{itemize}

Where applicable, we also validated correct identification of event types (e.g., Copy, Focus, Start) and ensured they appeared in the correct sequence. Table~\ref{tab:validation_metrics} summarizes the precision and recall metrics obtained during these tests. These results demonstrate \textit{CodeWatcher}'s robustness in capturing fine-grained user behaviors across a variety of realistic usage patterns.

\begin{table}
    \centering
    \footnotesize
    \caption{Validation Metrics Across Scenarios Summary}
    \label{tab:validation_metrics}
    \begin{tabular}{|p{2.35cm}|p{2.35cm}|p{1cm}|p{1cm}|}
    \hline
    \textbf{Scenario} & \textbf{Event Type(s)} & \textbf{Precision} & \textbf{Recall} \\
    \hline
    Insertion Accuracy & Insertion & 1.00 & 1.00 \\ \hline
    Deletion Detection & Deletion & 1.00 & 1.00 \\ \hline
    Copy-Paste Behavior & Copy, Paste & &  \\ \hline
    Focus Switching & Focus, Unfocus & 0.87 & 1.00 \\ \hline
    Back-and-Forth Use Patterns & Insertion, Deletion & 1.00 & 1.00 \\ \hline
    Multiple File Sessions & Start, Insertion (across tabs/file), end & 1.00 & 1.00 \\ \hline
    Event Logging Under Load (1 user) & All events & 0.93 & 1.00  \\ \hline
    Concurrent Users & All events (stress test) & 1.00 & 1.00 \\ \hline

    \end{tabular}
\end{table}





\subsection{Robustness Under Diverse Use Cases}

We also validated behavior under more realistic usage scenarios such as:

\begin{itemize}
    \item \textbf{Rapid Back-and-Forth Use Patterns:} Simulated high-speed editing and undo-redo operations.
   \item \textbf{Multi-line Suggestion Acceptance:} Accepted multiple lines from code suggestions to validate correct grouping and event generation (part of insertion validation).
  \item \textbf{Multiple File Sessions:} Utilized multiple files in parallel to test session consistency and file-scoped logging.
  \item \textbf{Concurrent Users:} Ran parallel sessions to ensure isolated logging and user separation with users.

\end{itemize}

\subsection{Summary of Findings}

Across all scenarios, on average, \textit{CodeWatcher} accurately detected and logged user interactions in accordance with the specification in Table~\ref{tab:event-types}. Captured data was successfully transmitted to the server, and timestamps and content fields aligned with observed behavior in the IDE. This validation confirms that \textit{CodeWatcher} reliably captures raw behavioral data necessary for downstream analyses, without relying on assumptions about specific analytics workflows or post-processing interpretations.

















\section{Empirical Use Case Example: Identifying AI Generated Code}

\begin{algorithm}[H]
\footnotesize
\caption{Algorithm for Categorizing Coding Interactions}
\label{ai_generated}
\KwIn{\textit{CodeWatcher} logs (JSON), Final submitted code (text file)}
\KwOut{Line-by-line labels: \textit{AI-Generated}, \textit{AI-Modified}, or \textit{User-Written}}

\BlankLine
\textbf{Step 1: Load and Preprocess Data} \\
Load the JSON file and extract historical code lines. \;
Read the final code file and split it into individual lines, normalize, and strip whitespace. \;

\BlankLine
\textbf{Step 2: Compare Lines and Compute Similarity} \\
\ForEach{line $L_f$ in final code}{
    \ForEach{line $L_h$ in historical data}{
        Compute fuzzy similarity score $S(L_f, L_h)$ \;
    }
    Let $L_{\text{best}}$ be the historical line with highest score \;
    \BlankLine

    \textbf{Step 3: Label Line Based on Similarity Score} \\
    \eIf{$S(L_f, L_{\text{best}}) \geq 95$}{
        Label $L_f$ as \textit{AI-Generated} \;
    }{
        \eIf{$80 \leq S(L_f, L_{\text{best}}) < 95$}{
            Label $L_f$ as \textit{AI-Modified} \;
        }{
            Label $L_f$ as \textit{User-Written} \;
        }
    }
}

\BlankLine
\textbf{Step 4: Save Output} \\
Store labeled lines into a structured JSON output file \;

\end{algorithm}

The \textit{CodeWatcher} tool can enable researchers to analyze the authorship of code submissions by linking keystroke-level interaction data with the final program output. To illustrate this, we implemented a post-processing script to be run in the server side, that determines whether each line of code in a final code submission was generated by a CGT, CGT generated but modified by the user, or entirely written by the user. This classification process begins with \textit{CodeWatcher}’s JSON log files, which record the code inserted from the CGT during the programming session. The final submitted code file is then compared to the historical logs (i.e., data from \textit{CodeWatcher} logs) using a fuzzy string matching algorithm (described in \ref{ai_generated}) that computes similarity scores between corresponding lines. Based on these scores, the script labels lines as \textit{AI-Generated} (score $\geq 95$), \textit{AI-Modified} (score between 80 and 94), or \textit{User-Written} (score $< 80$). This labeling allows researchers to quantify the extent of AI assistance, identify instances of user engagement or editing, and evaluate how students or developers interact with generative coding tools. The output can be further validated using precision, and recall, and F1-score metrics against manually annotated ground truth data. The methodology is reproducible, and adaptable across diverse datasets of coding activity.

The results on three developers completing easy tasks show on average, 68\% of the code written by participants was directly generated by an AI tool, 7\% was AI-generated but modified by the user, and 25\% was entirely user-written. The classification algorithm demonstrated high precision (77\%) but moderate recall (41.5\%). These results suggest that our use of \textit{CodeWatcher} data provides a promising foundation for reliably distinguishing between AI- and user-written code, though further improvements are needed to enhance sensitivity.

 
\section{Use Cases}

This section presents a set of use cases in which we envision \textit{CodeWatcher} being used.

\textbf{Educational Feedback and Assessment:} In academic settings instructors can use the data from \textit{CodeWatcher} to understand how students approach assignments beyond grading final code submissions. \textit{CodeWatcher} data can help identify when students overly on AI-generated completions or copy-paste code without meaningful engagement, supporting a better assessment of student effort and understanding. It can also inform curriculum refinement by revealing recurring pain points across peers.

\textbf{Research on Programming Behaviour:} \textit{CodeWatcher} offers researchers a robust platform for studying how novice and expert programmers interact with code, IDEs, and CGTs. It enables the collection of rich temporal and behavioral data, allowing for analyses of strategies such as incremental development, test-driven coding, or (over-)reliance on AI completions. Researchers can conduct controlled studies to compare coding behaviors across tools, environments, or demographics (e.g., native vs. non-native English speakers, gender groups) to assess how these factors influence learning and performance. Such data is critical for understanding equity in CS education and the role of AI in shaping different learners' experiences. 

\textbf{CGT Evaluation:} \textit{CodeWatcher} can be used to assess the utility, limitations, and impacts of CGTs on user interactions and outcomes. By logging all interactions with these tools such as when completions are accepted or edited, researchers can better explore where CGTs are helpful, confusing, or misused. These insights can inform refinements to AI suggestion engines, particularly in how suggestions are presented or triggered. \textit{CodeWatcher} data can also be used for retrospective attribution of which code segments were likely AI-generated versus hand-written, helping to disentangle human and machine contributions in collaborative development. 

\textbf{Industry Applications:} Development teams can use \textit{CodeWatcher} data to evaluate productivity based on interaction richness, and task-switching patterns. 
This can help identify blockers, improve tooling, or streamline workflows. For onboarding, \textit{CodeWatcher} allows teams to monitor how new developers learn internal systems and identify areas where additional support is needed. Engineering leads can also use \textit{CodeWatcher} to evaluate the adoption and real-world utility of new IDE features or plugins, identifying friction points or unused capabilities. In regulated industries or high-stakes collaborations, \textit{CodeWatcher} logs may serve as provenance data for audit trails, particularly when code authorship must be clarified.

\textbf{Ethics, Fairness, and AI Accountability:} \textit{CodeWatcher} contributes to responsible AI practices by enhancing transparency and traceability in code authored with CGTs. As AI-generated code becomes increasingly prevalent in both educational and industrial contexts, distinguishing between human and machine contributions is crucial for fair grading, authorship attribution, and intellectual property claims. 
It can also help identify potential biases in CGTs by revealing disparities in suggestion acceptance or effectiveness across user groups. \textit{CodeWatcher} empowers educators, researchers, and organizations to promote fairness, accountability, and informed decision-making in AI-assisted development.

\textbf{IDE Personalization and Smart Interventions:} Finally, \textit{CodeWatcher} lays the groundwork for intelligent, personalized IDEs. Real-time data on user interaction can be used to trigger context-sensitive interventions—such as hints, documentation links, or debugging suggestions such as when a user appears to be struggling. Over time, this data can also inform IDE adaptations that suggest shortcuts, identify common errors, or recommend tools tailored to individual developer workflows. By supporting these smart, adaptive features, \textit{CodeWatcher} can transform the static IDE into a more responsive and supportive learning and development environment.

\section{Limitations}

\textit{CodeWatcher} may not work as expected with different IDEs or CGTs, and does require a server for the user to connect to the tool so that the data is uploaded in real-time. However, the tool can be setup to run on a local server or locally without a server upload. 
Web-inserted code logs as unfocus and then insertion not paste, hinting it's not AI, but remains error-prone. The tool intentionally does not capture text that is not part of the coding language of the file. For example if a user inserts text that is not considered code, or a comment, the tool will not capture this as to avoid irrelevant data. This can make the plugin data less useful for those interested in such data.  

\section{Conclusion}

We presented \textit{CodeWatcher}, a lightweight plugin that captures coding interactions in the presence of CGTs. 
Our empirical use case shows how this data supports  classification of AI-generated, AI-modified, and user-written code. \textit{CodeWatcher} offers a practical foundation for studying CGT usage and developer behavior with minimal overhead.

\section*{Acknowledgments}
We thank GOOGLECA (PWPU GR028067) and CNPq (420406/2023-9 and 442779/2023-2) for funding this research.

\bibliographystyle{IEEEtran}
\bibliography{ref.bib}
\end{document}